\documentclass[preprint,eqsecnum,aps,nofootinbib]{revtex4}
\usepackage{amsfonts,amsmath,amssymb,amsthm}
\usepackage{latexsym}
\usepackage{bbm,bm}
\usepackage{graphicx}

\newcommand{\ket}[1]{\lvert #1 \rangle}
\newcommand{\bra}[1]{\langle #1 \lvert}
\newcommand{\beq}{\begin{equation}}
\newcommand{\eeq}{\end{equation}}
\newcommand{\beqs}{\begin{eqnarray}}
\newcommand{\eeqs}{\end{eqnarray}}
\begin{document}

\title{Difficulties in analytic computation for relative entropy of entanglement}

\author{Hungsoo Kim$^{1}$, Mi-Ra Hwang$^2$, Eylee Jung$^2$, DaeKil Park$^{2,3}$}

\affiliation{$^1$ The Institute of Basic Science, Kyungnam University, Masan, 631-701, Korea    \\
             $^2$ Department of Physics, Kyungnam University, Masan, 631-701, Korea \\
             $^3$ Department of Electronic Engineering, Kyungnam University, Masan, 631-701, Korea }

\begin{abstract}
It is known that relative entropy of entanglement for an entangled state $\rho$ is defined via its closest separable 
(or positive partial transpose) state $\sigma$. Recently, it has been shown how to find $\rho$ provided that $\sigma$ is given
in two-qubit system. In this paper we study on the reverse process-i.e., how to find $\sigma$ provided that $\rho$ is given. 
It is shown that if $\rho$ is one of Bell-diagonal, generalized Vedral-Plenio, and generalized Horodecki states, one can  
find $\sigma$ from a geometrical point of view. This is possible due to the following two facts:
(i) The Bloch vectors of $\rho$ and $\sigma$ are identical with each other (ii) The correlation vector of $\sigma$ 
can be computed from a crossing point between a minimal geometrical object, in which all separable states reside in the presence of 
Bloch vectors, and a straight line, which connects the point corresponding to the correlation vector of $\rho$ and 
the nearest vertex of the maximal tetrahedron, where all two-qubit states reside. It is shown, however, that these nice properties 
are not maintained for the arbitrary two-qubit states. 
\end{abstract}

\maketitle

\section{Introduction}
It is well known that entanglement of quantum states is an important physical resource in the context of the 
quantum information theories. It plays a crucial role in quantum teleportation\cite{bennett93}, superdense coding\cite{bennett92},
quantum cloning\cite{scarani05}, quantum cryptography\cite{ekert91} and quantum computer 
technology\cite{vidal03-1,nielsen00}\footnote{There are, however, several examples, where entanglement does not 
play an important role in the quantum computation. For example, the efficiency of Grover's search algorithm gets worsened if the initial 
state is entangled one\cite{shim04}. Another important example is a deterministic quantum computation with one pure
qubit\cite{lanyon08}. Other simple examples are presented in Ref.\cite{biham03}. Therefore, one cannot conclude definitely that entanglement is 
essential for quantum computation.}.
Therefore, to understand how to quantify and how to characterize the entanglement for a given quantum state is an highly important 
physical task. 

Many entanglement measures have been developed for last few years. 
Above all, in our opinion, the most important entanglement measure is a distillable entanglement\cite{bennett96}, which quantifies how
many maximally entangled states can be constructed from the copies of the given quantum state in the asymptotic region. The importance
of the distillable entanglement arises due to the fact that the entanglement is fragile when noises interfere the quantum information 
processing. The disadvantage of the distillable entanglement is its calculational difficulty. In order to compute the 
distillable entanglement analytically we should find the optimal purification (or distillation) protocol. If this optimal 
protocol generates $n$ maximally entangled states from $m$-copies of the quantum state $\rho$, the distillable entanglement for 
$\rho$ is given by\footnote{In Ref.\cite{bennett96} the distillable entanglement $D$ is divided into $D_1$ and $D_2$ depending 
on one-way and two-way classical communications. Throughout this paper we only consider the two-way classical 
communication.}
\begin{equation}
\label{distillable entanglement-1}
D(\rho) = \lim_{m \rightarrow \infty} \frac{n}{m}.
\end{equation}
However, finding an optimal purification protocol is an highly non-trivial task. It makes it difficult to compute the 
distillable entanglement analytically.
 
Fortunately, the tight upper bound of the distillable entanglement has been developed in Ref.\cite{vedral-97-1,vedral-97-2}.
In these references the new entanglement measure called relative entropy of entanglement (REE) was introduced. It is defined as 
\begin{equation}
\label{ree-1-1}
E_R (\rho) = \min_{\sigma \in {\cal D}} S(\rho || \sigma),
\end{equation}
where ${\cal D}$ is a set of separable states and $S(\rho || \sigma)$ is a quantum relative entropy-i.e.,
$S(\rho || \sigma) = \mbox{tr} (\rho \ln \rho - \rho \ln \sigma)$. It was shown in Ref.\cite{vedral-97-2} that $E_{R} (\rho)$ 
is a upper bound of the distillable entanglement. Subsequently, Rains\cite{rains-99-1,rains-99-2} has shown that 
\begin{equation}
\label{ree-1-2}
\tilde{E}_R (\rho) = \min_{\sigma \in {\cal D}_{PPT}} S(\rho || \sigma),
\end{equation}
where ${\cal D}_{PPT}$ is a set of positive partial transposition (PPT) states, is more tight upper bound when $\rho$ is an 
higher-dimensional bipartite state. Using the facts that the REE is an upper bound of the distillable entanglement and 
Smolin state\cite{smolin00} is a bound entangled state, the distillable entanglement for the various Bell-state mixtures
has been analytically computed\cite{ghosh01,chen02,chen03}. In order to understand the distillable entanglement more deeply, 
therefore, it is important to develop the various techniques for the explicit computation of the REE. 
Of course, regardless of the distillable entanglement, the development of the calculation technique for the REE itself is important
to understand the characterization of entanglement more profoundly. 
For last few years many properties of the REE were investigated\cite{horo-07-1}. 
Furthermore, recently relation between the REE and other distance measures has been studied\cite{hayashi05,wei08}.

In this paper we confine ourselves to the REE when $\rho$ is two-qubit states-i.e., 
$\rho \in {\cal H}^2 \otimes {\cal H}^2$. Since there is no bound-entangled state in this case, $E_R(\rho)$ and 
$\tilde{E}_R (\rho)$ defined in Eq.(\ref{ree-1-1}) and Eq.(\ref{ree-1-2}) are same. Let $\sigma^*$ be 
the closest separable state (CSS) of $\rho$. Then, $E_R(\rho)$ is given by
\begin{equation}
\label{ree-1-3}
E_R (\rho) = \min_{\sigma \in {\cal D}} S(\rho || \sigma) = S(\rho || \sigma^*).
\end{equation}
When the CSS $\sigma^*$ is explicitly given and it is full rank, Ref.\cite{miran-08-1} has presented how to construct the set of the
entangled states, whose CSS are $\sigma^*$. Let $\ket{i}$ and $\lambda_i$ be eigenvectors and corresponding eigenvalues 
of $\sigma^*$. If $\sigma^*$ is the CSS (hence, edge) state, then its partial transposition $\sigma^{\Gamma}$ is rank deficient. 
Let $\ket{\phi}$ be the kernel of $\sigma^{\Gamma}$-i.e.,
\begin{equation}
\label{kernel-1-1}
\sigma^{\Gamma} \ket{\phi} = 0.
\end{equation}
Then, the set of the entangled states $\rho(x)$, whose CSS are $\sigma^*$, is given by the following one-parameter family
expression:
\begin{eqnarray}
\label{one-parameter-1-1}
& & \rho(x) = \sigma^* - x G(\sigma^*)   \\  \nonumber
& & G(\sigma^*) = \sum_{i,j} G_{ij} \ket{i} \bra{i} (\ket{\phi} \bra{\phi})^{\Gamma} \ket{j} \bra{j},
\end{eqnarray}
where $x \geq 0$ and 
\begin{eqnarray}
\label{bozo-1-1}
G_{ij} = \left\{    \begin{array}{cc}
                \lambda_i  &   \hspace{.5cm} \mbox{for} \hspace{.2cm} i=j    \\
                \frac{\lambda_i - \lambda_j}{\ln \lambda_i - \ln \lambda_j}   & \hspace{.5cm}  \mbox{for} \hspace{.2cm} i \neq j.
                    \end{array}          \right.
\end{eqnarray}					
When, however, the entangled state $\rho$ is explicitly given, it is difficult to use Eq.(\ref{one-parameter-1-1}) for 
finding its CSS. In other words we have to find the reverse process of Ref.\cite{miran-08-1} in order to derive the 
closed formula of the REE for the arbitrary two-qubit state $\rho$ as Wootters\cite{woot-98} has done in the entanglement of formation. 
Unfortunately, still it is an unsolved problem\cite{open05}.

In this paper we will explore the reverse process of Ref.\cite{miran-08-1}. We will show that the reverse process
of Ref.\cite{miran-08-1} is possible, at least, for the Bell-diagonal, generalized Vedral-Plenio, and generalized 
Horodecki states. We present a method for finding the corresponding CSS systematically for these states by generalizing the geometrical 
method discussed by Horodecki in Ref.\cite{horo-96-1}. We also discuss why it is difficult to find the CSS for the 
arbitrary two-qubit mixed states from the geometrical point of view. The paper is organized as follows.
In section II we will show how to find the CSS for the Bell-diagonal states. In section III we discuss how the geometrical
objects presented in Ref.\cite{horo-96-1} such as tetrahedron ${\cal T}$ and octahedron ${\cal L}$ 
are deformed in the presence of the 
non-zero Bloch vectors. In section IV and section V we present a method for finding the corresponding CSS for the generalized 
Vedral-Plenio and generalized Horodecki states, respectively. In section VI we discuss why it is very difficult task to find the CSS for the 
arbitrary two-qubit states from the geometrical point of view. In section VII a brief conclusion is given.

\section{CSS for the Bell-Diagonal States}
In this section we show how to find the CSS when $\rho$ is the Bell-diagonal state from the geometrical point 
of view. In fact, this problem was already solved in Ref.\cite{vedral-97-2} long ago. The reason why we re-consider the same 
problem is to stress the geometrical analysis.

An arbitrary two-qubit state can be represented as follows:
\begin{equation}
\label{arbitrary-2-1}
\rho = \frac{1}{4} \left[ I \otimes I + {\bm r} \cdot {\bm \sigma} \otimes I + I \otimes {\bm s} \cdot {\bm \sigma} + 
                           \sum_{m,n = 1}^3 g_{mn} \sigma_m \otimes \sigma_n \right],
\end{equation}
where ${\bm r}$ and ${\bm s}$ are Bloch vectors and $\sigma_i$ is usual Pauli matrix. The coefficients $g_{mn}$ form a 
real matrix and represent the interaction of the qubits. If state $\rho$ is explicitly given, one can derive the 
Bloch vectors ${\bm r}$ and ${\bm s}$ and the correlation tensor $g_{ij}$ as follows:
\begin{equation}
\label{bozo-2-1}
{\bm r} = \mbox{tr} (\rho_A {\bm \sigma}),    \hspace{1.0cm}
{\bm s} = \mbox{tr} (\rho_B {\bm \sigma}),    \hspace{1.0cm}
g_{ij} = \mbox{tr} (\rho \sigma_i \otimes \sigma_j),
\end{equation}
where $\rho_A = \mbox{tr}_B \rho$ and $\rho_B = \mbox{tr}_A \rho$.
It is well known that an appropriate local unitary (LU) transformation of $\rho$ can make $g_{mn}$ to be diagonal 
(see appendix of Ref.\cite{jung-08-1}). Since entanglement is invariant under the LU transformation, it is in general 
sufficient to consider the case of diagonal $g_{mn}$ for the discussion of entanglement. Thus, without loss of generality,
one can express $\rho$ as 
\begin{equation}
\label{arbitrary-2-2}
\rho = \frac{1}{4} \left[ I \otimes I + {\bm r} \cdot {\bm \sigma} \otimes I + I \otimes {\bm s} \cdot {\bm \sigma} + 
                           \sum_{n = 1}^3 g_{n} \sigma_n \otimes \sigma_n \right].
\end{equation}
If $\rho = \ket{\beta_i} \bra{\beta_i}$, where
\begin{eqnarray}
\label{bell-2-1}
& &\ket{\beta_1} = \frac{1}{\sqrt{2}} \left( \ket{00} + \ket{11} \right)     \hspace{2.0cm}
   \ket{\beta_2} = \frac{1}{\sqrt{2}} \left( \ket{00} - \ket{11} \right)     \\   \nonumber
& &\ket{\beta_3} = \frac{1}{\sqrt{2}} \left( \ket{01} + \ket{10} \right)     \hspace{2.0cm}
   \ket{\beta_4} = \frac{1}{\sqrt{2}} \left( \ket{01} - \ket{10} \right) ,
\end{eqnarray}
it is easy to show that the corresponding Bloch vectors ${\bm r}$ and ${\bm s}$ are vanishing and the corresponding
correlation tensor $g_{mn}$ become
\begin{equation}
\label{bell-2-2}
g_1 = \mbox{diag}(1,-1,1)  \hspace{.5cm} g_2 = \mbox{diag}(-1,1,1) \hspace{.5cm} g_3 = \mbox{diag}(1,1,-1) 
\hspace{.5cm} g_4 = \mbox{diag}(-1,-1,-1).
\end{equation}
If, therefore, $\rho$ is Bell-diagonal state, the Bloch vectors ${\bm r}$ and ${\bm s}$ are always null vectors.

Since we are considering on the diagonal case of the correlation tensor, we will regard, from now on, the tensor as a 
vector, whose components are equal to the diagonal elements. When ${\bm r} = {\bm s} = 0$, Horodecki has shown in 
Ref.\cite{horo-96-1} that the total two-qubit states belong to the tetrahedron ${\cal T}$ with vertices 
$v_1 = (1,-1,1)$, $v_2 = (-1,1,1)$, $v_3 = (1,1,-1)$ and $v_4 = (-1,-1,-1)$ in the correlation vector space.
Ref.\cite{horo-96-1} also has shown that the separable states (with ${\bm r} = {\bm s} = 0$) belong to the 
octahedron ${\cal L}$ with vertices $o_1^{(\pm)} = (\pm 1, 0, 0)$, $o_2^{(\pm)} = (0, \pm 1, 0)$ and 
$o_3^{(\pm)} = (0, 0, \pm 1)$. This is pictorially represented in Fig. 1.

%%%%%%%%%%%%%%%%%%%%%%%%%%%%%%%%%%%%%%%%%%%%%%%%%%%%%%%%%
\begin{figure}[ht!]
\begin{center}
\caption[fig1]{The total Bell-diagonal states belong to the tetrahedron $(v_1, v_2, v_3, v_4)$ and the set of
the separable states belong to the octahedron, whose vertices are $o_1^{\pm}$, $o_2^{\pm}$ and $o_3^{\pm}$.  
As this figure shows, the planes $(o_1^{(+)}, o_2^{(-)}, o_3^{(-)})$, $(o_1^{(+)}, o_2^{(+)}, o_3^{(+)})$, 
$(o_1^{(-)}, o_2^{(-)}, o_3^{(+)})$ and $(o_1^{(-)}, o_2^{(+)}, o_3^{(-)})$ are contained in the planes
$(v_1, v_3, v_4)$, $(v_1, v_2, v_3)$, $(v_1, v_2, v_4)$ and $(v_2, v_3, v_4)$, respectively. Therefore, all 
entangled Bell-diagonal mixtures belong to the small four tetrahedra $(v_1, o_1^{(+)}, o_2^{(-)}, o_3^{(+)})$, 
$(v_2, o_1^{(-)}, o_2^{(+)}, o_3^{(+)})$, $(v_3, o_1^{(+)}, o_2^{(+)}, o_3^{(-)})$ and 
$(v_4, o_1^{(-)}, o_2^{(-)}, o_3^{(-)})$.}
\end{center}
\end{figure}
%%%%%%%%%%%%%%%%%%%%%%%%%%%%%%%%%%%%%%%%%%%%%%%%%%%%%%%%%%%

As Fig. 1 shows, the planes $(o_1^{(+)}, o_2^{(-)}, o_3^{(-)})$, $(o_1^{(+)}, o_2^{(+)}, o_3^{(+)})$, 
$(o_1^{(-)}, o_2^{(-)}, o_3^{(+)})$ and $(o_1^{(-)}, o_2^{(+)}, o_3^{(-)})$ are parts of the planes
$(v_1, v_3, v_4)$, $(v_1, v_2, v_3)$, $(v_1, v_2, v_4)$ and $(v_2, v_3, v_4)$ ,respectively\footnote{This statement 
can be confirmed by deriving the respective plane equations. The plane equations for 
$(v_1, v_3, v_4)$, $(v_1, v_2, v_3)$, $(v_1, v_2, v_4)$ and $(v_2, v_3, v_4)$ are $x - y - z=1$, 
$x + y + z=1$, $-x - y + z=1$ and $-x + y - z=1$, respectively. It is easy to show that these plane equations
are the same planes with the planes $(o_1^{(+)}, o_2^{(-)}, o_3^{(-)})$, $(o_1^{(+)}, o_2^{(+)}, o_3^{(+)})$, 
$(o_1^{(-)}, o_2^{(-)}, o_3^{(+)})$ and $(o_1^{(-)}, o_2^{(+)}, o_3^{(-)})$, respectively.}. Therefore, all entangled 
Bell-diagonal mixtures belong to the small four tetrahedra $(v_1, o_1^{(+)}, o_2^{(-)}, o_3^{(+)})$, 
$(v_2, o_1^{(-)}, o_2^{(+)}, o_3^{(+)})$, $(v_3, o_1^{(+)}, o_2^{(+)}, o_3^{(-)})$ and 
$(v_4, o_1^{(-)}, o_2^{(-)}, o_3^{(-)})$.

Now, we show how to perform the reverse process of Ref.\cite{miran-08-1} when $\rho$ is an 
entangled Bell-diagonal state. This can be achieved by following two theorems.

{\bf Theorem 1.} {\it Every Bell state has infinite CSS, which cover fully the nearest surface of the octahedron ${\cal L}$.} 

\smallskip

{\bf Proof.} It is sufficient to prove this theorem when $\rho = \ket{\beta_1} \bra{\beta_1}$. When 
$\rho = \ket{\beta_i} \bra{\beta_i} \hspace{.1cm} (i=2, 3, 4)$, one can prove the theorem similarly.

Let $\sigma$ be a following Bell-diagonal state:
\begin{equation}
\label{theorem-1-1}
\sigma = \frac{1}{4} \left[ I \otimes I + \sum_{n=1}^3 p_n \sigma_n \otimes \sigma_n \right]
\end{equation}
with ${\bm p} = (x, y, z)$. Then, it is easy to show that the spectral decomposition of $\sigma$ is 
\begin{eqnarray}
\label{theorem-1-2}
& &\sigma = \frac{1+x-y+z}{4} \ket{\beta_1} \bra{\beta_1} + \frac{1-x+y+z}{4} \ket{\beta_2} \bra{\beta_2} \\  \nonumber
& & \hspace{.7cm} + \frac{1+x+y-z}{4} \ket{\beta_3} \bra{\beta_3} + \frac{1-x-y-z}{4} \ket{\beta_4} \bra{\beta_4}.
\end{eqnarray}
The nearest surface of ${\cal L}$ from $\rho = \ket{\beta_1} \bra{\beta_1}$ is $(o_1^{(+)},o_2^{(-)},o_3^{(+)})$, 
whose surface equation is $x-y+z=1$. If $\sigma$ belongs to the surface $(o_1^{(+)},o_2^{(-)},o_3^{(+)})$, it is 
easy to show that $S(\rho || \sigma) = \ln 2$, which exactly coincides with the REE of the Bell states\cite{vedral-97-1}.
Therefore, $\sigma$ on the surface $(o_1^{(+)},o_2^{(-)},o_3^{(+)})$ is the CSS of $\ket{\beta_1} \bra{\beta_1}$. 

Now, let us consider the case where $\sigma$ belongs to other surface. For example, let us assume that $\sigma$ 
belongs to the surface $(o_1^{(+)},o_2^{(+)},o_3^{(-)})$, whose surface equation is $x+y-z =1$. Then, 
$S(\rho || \sigma)$ reduces to $\ln 2 - \ln x$, which is less than $\ln 2$ if $x \neq 1$. Therefore $\sigma$ on 
$(o_1^{(+)},o_2^{(+)},o_3^{(-)})$ is not CSS of $\ket{\beta_1} \bra{\beta_1}$. By same way one can show that 
$\sigma$ on $(o_1^{(-)},o_2^{(+)},o_3^{(+)})$ or $(o_1^{(-)},o_2^{(-)},o_3^{(-)})$ is not CSS of 
$\ket{\beta_1} \bra{\beta_1}$, which completes the proof.

\smallskip

{\bf Theorem 2.} {\it The CSS of the any Bell-diagonal state $\rho$ corresponds to the crossing point between
the nearest surface of ${\cal L}$ from $\rho$ and the straight line $\ell$, which connects 
$\rho$ and the nearest vertex of ${\cal T}$ from $\rho$.} 

\smallskip

{\bf Proof.} If $\sigma$ is CSS of $\rho$, the CSS of $\tilde{\rho} = x \rho + (1-x) \sigma$ is also 
$\sigma$\cite{vedral-97-2}. Let $\rho$ be $\rho = \ket{\beta_1} \bra{\beta_1}$. Then, theorem 1 implies that 
$\sigma$ can be any point on the surface $(o_1^{(+)},o_2^{(-)},o_3^{(+)})$. 
Let $\tilde{\rho}$ belong to the small tetrahedron
$(v_1, o_1^{(+)},o_2^{(-)},o_3^{(+)})$. Note that $\tilde{\rho}$ corresponds to a internally dividing 
point of the line segment $\overline{\rho \sigma}$. Since Eq.(\ref{one-parameter-1-1}) implies that the set of the 
entangled states which have same CSS should be represented by the straight line, the only possible $\sigma$ as 
CSS of $\tilde{\rho}$ is a
crossing point between a line $\overline{\rho \tilde{\rho}}$ and the surface $(o_1^{(+)},o_2^{(-)},o_3^{(+)})$,
which completes the proof for the Bell-diagonal states.

\smallskip

%%%%%%%%%%%%%%%%%%%%%%%%%%%%%%%%%%%%%%%%%%%%%%%%%%%%%%%%%
\begin{figure}[ht!]
\begin{center}
\includegraphics[height=10cm]{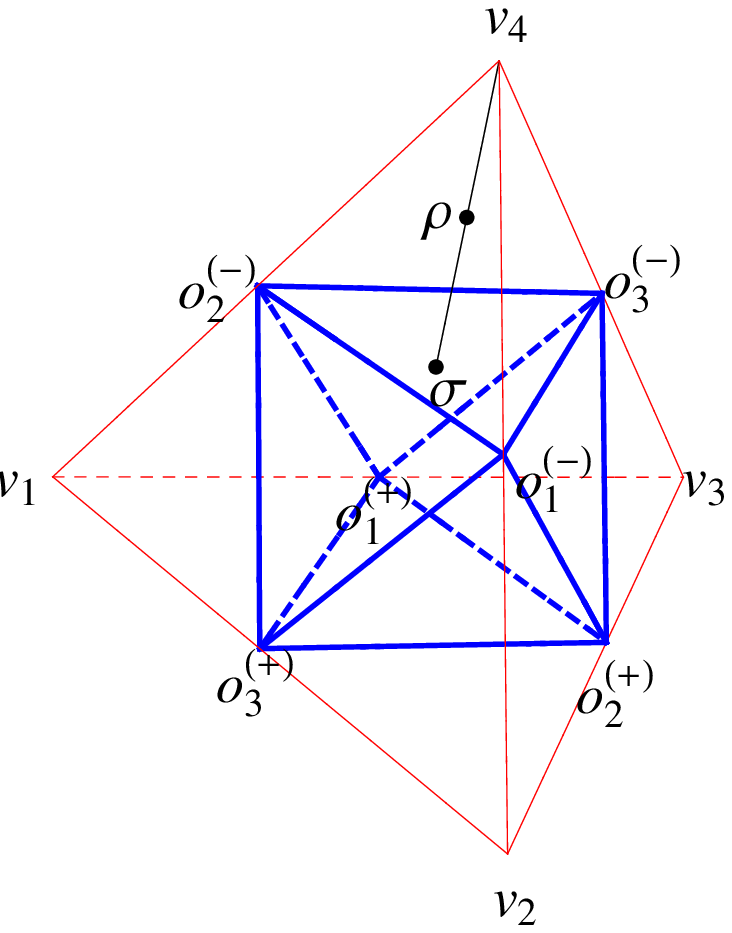}
\caption[fig2]{This figure shows how to find the CSS for the Bell-diagonal state. First, extend the line segment between 
$\rho$ and 
the point corresponding to the nearest vertex of ${\cal T}$. Second, compute the coordinate of the crossing point 
between the line and the nearest surface of the octahedron ${\cal L}$. Finally, find the CSS of $\rho$ which corresponds to the 
crossing point.}
\end{center}
\end{figure}
%%%%%%%%%%%%%%%%%%%%%%%%%%%%%%%%%%%%%%%%%%%%%%%%%%%%%%%%%%%

By making use of the Theorem 2 one can always find the CSS $\sigma$ if $\rho$ is a Bell-diagonal state. 
Fig. 2 shows how to find the CSS for the Bell-diagonal state. First, extend the line segment between $\rho$ and 
the point corresponding to the nearest vertex of ${\cal T}$. Second, compute the coordinate of the crossing point 
between the line and the nearest surface of the octahedron ${\cal L}$. Finally, find the CSS which corresponds to the 
crossing point. This complete the reverse process of Ref.\cite{miran-08-1}.

\section{Geometrical Deformation of ${\cal T}$ and ${\cal L}$}
When the Bloch vectors ${\bm r}$ and ${\bm s}$ are non-zero, the geometrical objects ${\cal T}$ and ${\cal L}$ 
should be deformed. In this section we will discuss how ${\cal T}$ and ${\cal L}$ are deformed. In order to perform 
the following analysis analytically we consider in this paper the case where ${\bm r}$ and ${\bm s}$ are parallel to each other. 
It is worthwhile noting that if ${\bm r}$ and ${\bm s}$ are $x$- or $y$-direction, one can make them to be $z$-directional via the appropriate 
local-unitary transformation. For example, if they are $x$-direction, $\rho' = (U \otimes U) \rho (U \otimes U)^{\dagger}$ with
$$ U = \frac{1}{\sqrt{2}} \left(         \begin{array}{cc}
                                           1  &  1  \\
										   -1  &  1 
										   \end{array}         \right)    $$
has $z$-directional Bloch vectors and its correlation vector changes from $(g_1, g_2, g_3)$ to $(-g_3, g_2, g_1)$. 
Similarly, one can change the state with $y$-directional Bloch vectors into the state with $z$-directional Bloch vectors without 
altering the diagonal property of the correlation term.

In this reason it is reasonable to assume that the directions of the Bloch vectors 
are $z$-direction by writing ${\bm r} = (0, 0, r)$ and ${\bm s} = (0, 0, s)$\footnote{Even if $\bm{r}$ and $\bm{s}$ are not parallel
with each other, one can make them to be $z$-directional via an appropriate local-unitary transformation. In this case, however,
the correlation term loses its diagonal property.}. In this case the arbitrary 
two-qubit state $\rho$ defined in Eq.(\ref{arbitrary-2-2}) with $\bm{g} = (q_1, q_2, q_3)$ reduces to
\begin{eqnarray}
\label{special-1}
\rho = \frac{1}{4} \left(            \begin{array}{cccc}
                          1+r+s+q_3  &  0  &  0  &  q_1 - q_2   \\
						  0  &  1+r-s-q_3  &  q_1 + q_2  &  0   \\
						  0  &  q_1 + q_2  &  1-r+s-q_3  &  0   \\
						  q_1 - q_2  &  0  &  0  &  1-r-s+q_3
						              \end{array}               \right).
\end{eqnarray}

%%%%%%%%%%%%%%%%%%%%%%%%%%%%%%%%%%%%%%%%%%%%%%%%%%%%%%%%%
\begin{figure}[ht!]
\begin{center}
%\includegraphics[height=8cm]{fig3a.eps}
%\hspace{1.0cm}
%\includegraphics[height=8cm]{fig3b.eps}
%\includegraphics[height=8cm]{fig3c.eps}
%\hspace{1.0cm}
%\includegraphics[height=8cm]{fig3d.eps}
\caption[fig3]{The deformation of ${\cal T}$ is plotted when $r=s=0.3$ (Fig. 3a), $r=-s=0.3$ (Fig. 3b), 
$r=s=0.5$ (Fig. 3c) and $r=-s=0.5$ (Fig. 3d). For comparison we plot ${\cal T}$ together. 
The appearance of non-zero Bloch vectors generally shrink the 
tetrahedron. The shrinking rate becomes larger with increasing the norm of the Bloch vectors.}
\end{center}
\end{figure}
%%%%%%%%%%%%%%%%%%%%%%%%%%%%%%%%%%%%%%%%%%%%%%%%%%%%%%%%%%%

The eigenvalues and eigenvectors of $\rho$ are summarized at Table I.

\begin{center}
\begin{tabular}{c|c} 
eigenvalues of $\rho$ & eigenvectors of $\rho$                                     \\  \hline \hline
$\mu_{\pm} = \frac{1}{4} \left\{ (1 - q_3) \pm M_1 \right\}$ &  
$\ket{\mu_{\pm}} = \frac{1}{\Lambda_{\pm}} \left[(q_1 + q_2) \ket{01} - \left\{ (r-s) \mp M_1 \right\} \ket{10} \right]$  \\    
$\nu_{\pm} = \frac{1}{4} \left\{ (1 + q_3) \pm M_2 \right\}$ &  
$\ket{\nu_{\pm}} = \frac{1}{\Delta_{\pm}} \left[(q_1 - q_2) \ket{00} - \left\{ (r+s) \mp M_2 \right\} \ket{11} \right]$ 
\end{tabular}

\vspace{0.3cm}
Table I: Eigenvalues and Eigenvectors of $\rho$ in Eq.(\ref{special-1})
\end{center}
\vspace{0.5cm}

At Table I $M_1$, $M_2$, $\Lambda_{\pm}$ and $\Delta_{\pm}$ are given by
\begin{eqnarray}
\label{bozo-3-1}
& &M_1 = \sqrt{(r-s)^2 + (q_1 + q_2)^2}  \hspace{1.0cm} M_2 = \sqrt{(r+s)^2 + (q_1 - q_2)^2}   \\  \nonumber
& &\Lambda_{\pm} = \sqrt{ \left\{(r-s) \mp M_1\right\}^2 + (q_1 + q_2)^2} \hspace{.5cm}
   \Delta_{\pm} = \sqrt{ \left\{(r+s) \mp M_2\right\}^2 + (q_1 - q_2)^2}.
\end{eqnarray}
Then, the deformation of ${\cal T}$ can be obtained from the positivity condition of $\rho$. Since deformation 
should be a set of the boundary states, the condition of the deformation becomes
\begin{equation}
\label{deform-1}
\min (\mu_-, \nu_-) = 0.
\end{equation}
One can make two surfaces by making use of Eq.(\ref{deform-1}). Each surface corresponds to 
$\min (\mu_-, \nu_-) = \mu_- = 0$ or $\min (\mu_-, \nu_-) = \nu_- = 0$. Gluing these surfaces together yields the deformation of 
${\cal T}$. 

In Fig. 3 we plot the deformation of ${\cal T}$ when $r=s=0.3$ (Fig. 3a), $r=-s=0.3$ (Fig. 3b), $r=s=0.5$ (Fig. 3c) and
$r=-s=0.5$ (Fig. 3d). For comparison we plot ${\cal T}$ together. For convenience, we will call the deformation of 
${\cal T}$ with fixed $r$ and $s$ as ${\cal T}_{r,s}$. From Fig. 3 one can realize that the deformation ${\cal T}_{r,s}$
has following two characteristics. First one is that the effect of the non-zero Bloch vectors is to shrink the geometrical object. 
The shrinking rate becomes larger with increasing $|r|$ and $|s|$. When $r=s$, the deformation is biased toward $(v_1, v_2)$ region. 
When, however, $r=-s$, the deformation is biased toward $(v_3, v_4)$ region. The shrinkage of ${\cal T}_{r,s}$ implies that 
the number of proper quantum states reduces with increasing $|r|$ and $|s|$ due to the constraint $\mbox{tr} \rho^2 \leq 1$.
The second characteristic of ${\cal T}_{r,s}$ is that it has continuous smooth surface while ${\cal T}$ has sharp edges.
This fact arises from the condition $\min (\mu_- , \nu_-) = 0$. When $r=s=0$, this condition generates the four surface equations
\begin{equation}
\label{star-1}
\pm (q_1 + q_2) + q_3 = 1    \hspace{1.0cm}  \pm (q_1 - q_2) - q_3 = 1,
\end{equation}
each of which corresponds to the surface of ${\cal T}$. When, however, $r$ and $s$ are non-zero, these four equations reduce
to the following two equations:
\begin{equation}
\label{star-2}
\sqrt{(r-s)^2 + (q_1 + q_2)^2} + q_3 = 1   \hspace{1.0cm} \sqrt{(r+s)^2 + (q_1 - q_2)^2} - q_3 = 1.
\end{equation}
This implies that the deformation ${\cal T}_{r,s}$ can be formed by attaching two smooth surfaces when $r \neq \pm s$.

%%%%%%%%%%%%%%%%%%%%%%%%%%%%%%%%%%%%%%%%%%%%%%%%%%%%%%%%%
\begin{figure}[ht!]
\begin{center}
%\includegraphics[height=8cm]{fig4a.eps}
%\hspace{1.0cm}
%\includegraphics[height=8cm]{fig4b.eps}
%
%\includegraphics[height=8cm]{fig4c.eps}
%\hspace{1.0cm}
%\includegraphics[height=8cm]{fig4d.eps}
\caption[fig4]{The deformation of ${\cal L}$ is plotted when $r=s=0.3$ (Fig. 4a), $r=-s=0.3$ (Fig. 4b), 
$r=s=0.5$ (Fig. 4c) and $r=-s=0.5$ (Fig. 4d). For comparison we plot ${\cal L}$ together. 
The appearance of non-zero Bloch vectors generally shrinks the 
octahedron. The shrinking rate becomes larger with increasing the norm of the Bloch vectors}
\end{center}
\end{figure}
%%%%%%%%%%%%%%%%%%%%%%%%%%%%%%%%%%%%%%%%%%%%%%%%%%%%%%%%%%%

Now, we discuss the deformation of ${\cal L}$ when the Bloch vectors are ${\bm r} = (0,0,r)$ and 
${\bm s} = (0,0,s)$. We will call this deformation as ${\cal L}_{r,s}$. 
We assume that $\rho$ in Eq.(\ref{special-1}) is a separable state. In this case the PPT state of $\rho$, 
{\it say} $\rho^{\Gamma}$,
should be positive. The eigenvalues and the corresponding eigenvectors of $\rho^{\Gamma}$ are summarized in Table II.

\begin{center}
\begin{tabular}{c|c} 
eigenvalues of $\rho^{\Gamma}$ & eigenvectors of $\rho^{\Gamma}$                                     \\  \hline \hline
$\mu_{\pm}^{\Gamma} = \frac{1}{4} \left\{ (1 - q_3) \pm M_1^{\Gamma} \right\}$ &  
$\ket{\mu_{\pm}^{\Gamma}} = \frac{1}{\Lambda_{\pm}^{\Gamma}} \left[(q_1 - q_2) \ket{01} - 
                                            \left\{ (r-s) \mp M_1^{\Gamma} \right\} \ket{10} \right]$  \\    
$\nu_{\pm}^{\Gamma} = \frac{1}{4} \left\{ (1 + q_3) \pm M_2^{\Gamma} \right\}$ &  
$\ket{\nu_{\pm}^{\Gamma}} = \frac{1}{\Delta_{\pm}^{\Gamma}} \left[(q_1 + q_2) \ket{00} - 
                                            \left\{ (r+s) \mp M_2^{\Gamma} \right\} \ket{11} \right]$ 
\end{tabular}

\vspace{0.3cm}
Table II: Eigenvalues and Eigenvectors of $\rho^{\Gamma}$
\end{center}
\vspace{0.5cm}

At Table II $M_1^{\Gamma}$, $M_2^{\Gamma}$, $\Lambda_{\pm}^{\Gamma}$ and $\Delta_{\pm}^{\Gamma}$ are defined as
\begin{eqnarray}
\label{bozo-3-2}
& & \hspace{2.0cm} M_1^{\Gamma} = M_1 \bigg|_{q_2 \rightarrow -q_2} = \sqrt{(r-s)^2 + (q_1 - q_2)^2}   \\  \nonumber
& & \hspace{2.0cm} M_2^{\Gamma} = M_2 \bigg|_{q_2 \rightarrow -q_2} = \sqrt{(r+s)^2 + (q_1 + q_2)^2}   \\  \nonumber
& &\Lambda_{\pm}^{\Gamma} = \sqrt{ \left\{(r-s) \mp M_1^{\Gamma} \right\}^2 + (q_1 - q_2)^2} \hspace{.5cm}
   \Delta_{\pm}^{\Gamma} = \sqrt{ \left\{(r+s) \mp M_2^{\Gamma} \right\}^2 + (q_1 + q_2)^2}.
\end{eqnarray}
Then ${\cal L}_{r,s}$ can be obtained from the positivity condition of $\rho^{\Gamma}$. Since, furthermore, 
${\cal L}_{r,s}$ should be a set of the edge states, the condition for the deformation of ${\cal L}$ reduces to 
\begin{equation}
\label{deform-2}
\min \left(\mu_-^{\Gamma}, \nu_-^{\Gamma} \right) = 0.
\end{equation}
As deformation of ${\cal T}$ Eq.(\ref{deform-2}) generates two surfaces, each of which corresponds to 
$\min \left(\mu_-^{\Gamma}, \nu_-^{\Gamma} \right) = \mu_-^{\Gamma} = 0$ or 
$\min \left(\mu_-^{\Gamma}, \nu_-^{\Gamma} \right) = \nu_-^{\Gamma} = 0$. Gluing these two surfaces one can make the 
deformation of ${\cal L}$.

In Fig. 4 we plot the deformation of ${\cal L}$ at $r=s=0.3$ (Fig. 4a), $r=-s=0.3$ (Fig. 4b), $r=s=0.5$ (Fig. 4c)
and $r=-s=0.5$ (Fig. 4d). For comparison we plot ${\cal L}$ together. Like the deformation of ${\cal T}$,
Fig. 4 indicates that the effect 
of the non-zero Bloch vectors is to shrink ${\cal L}$ toward a particular direction. Fig. 4 also shows that the 
shrinking rate becomes larger and larger with increasing the norm of the Bloch vectors. Like ${\cal T}_{r,s}$ again, 
the deformation of the octahedron ${\cal L}_{r,s}$ also has smooth surfaces while ${\cal L}$ has sharp edges.

\section{CSS for The Generalized Vedral-Plenio States}
In this section we show how to derive the CSS for the Vedral-Plenio (VP) states. The VP states are defined 
as mixture of one Bell state and separable states, which are not orthogonal to the Bell state. One but most general 
example of the VP state is 
\begin{equation}
\label{VP-1}
\rho_{vp} = \lambda_1 \ket{\beta_1} \bra{\beta_1} + \lambda_2 \ket{00} \bra{00} + \lambda_3 \ket{11} \bra{11},
\end{equation}
where $\ket{\beta_1} = (1 / \sqrt{2}) (\ket{00} + \ket{11})$ and $\lambda_1 + \lambda_2 + \lambda_3 = 1$. 

Let the arbitrary VP state be 
\begin{equation}
\label{VP-2}
\rho_{vp} = \frac{1}{4} \left( I \otimes I + {\bm r}_{vp} \cdot {\bm \sigma} \otimes I + 
I \otimes {\bm s}_{vp} \cdot {\bm \sigma} + \sum_{n=1}^3 ({\bm t}_{vp})_n \sigma_n \otimes \sigma_n \right)
\end{equation}
and its CSS be 
\begin{equation}
\label{VP-3}
\pi_{vp} = \frac{1}{4} \left( I \otimes I + {\bm u}_{vp} \cdot {\bm \sigma} \otimes I + 
I \otimes {\bm v}_{vp} \cdot {\bm \sigma} + \sum_{n=1}^3 ({\bm \tau}_{vp})_n \sigma_n \otimes \sigma_n \right).
\end{equation}
The following theorem shows how to compute ${\bm u}_{vp}$, ${\bm v}_{vp}$ and ${\bm \tau}_{vp}$
from $\rho_{vp}$. 

{\bf Theorem 3.} {\it If $\pi_{vp}$ is the CSS of $\rho_{vp}$, ${\bm u}_{vp} = {\bm r}_{vp}$ and 
${\bm v}_{vp} = {\bm s}_{vp}$. Let $\ell$ be a straight line, which connects ${\bm t}_{vp}$ and the 
nearest vertex of ${\cal T}$. Then, ${\bm \tau}_{vp}$ is a crossing point between $\ell$ and ${\cal L}_{r,s}$.} 

\smallskip

{\bf Proof.} We will prove this theorem as following procedure. First, we assume that this theorem is correct.
Then, following this theorem one can derive the trial CSS state of $\rho_{vp}$. Next, by making use of 
Eq.(\ref{one-parameter-1-1}) we will show that this trial CSS state is a really CSS state of $\rho_{vp}$.

Since other VP states can be derived from Eq.(\ref{VP-1}) by local-unitary (LU) transformation,
it is sufficient to show that the CSS of $\rho_{vp}$ in Eq.(\ref{VP-1}) satisfies this theorem. The other case 
can be proven similarly.
For $\rho_{vp}$ in Eq.(\ref{VP-1}) ${\bm r}_{vp}$, ${\bm s}_{vp}$ and ${\bm t}_{vp}$ become ${\bm r}_{vp} = (0,0,r)$, 
${\bm s}_{vp} = (0,0,s)$ and ${\bm t}_{vp} = (t_1, t_2, t_3)$ where
\begin{equation}
\label{theorem-3-1}
r = s = \lambda_2 - \lambda_3  \hspace{1.0cm} t_1 = -t_2 = \lambda_1  \hspace{1.0cm} t_3 = 1.
\end{equation}
Then, a point $P = (q_1, q_2, q_3)$ on the line $\ell$ satisfies $q_2 = -q_1$ and $q_3=1$. If the point 
$P = (q_1, q_2, q_3)$ is a crossing point between $\ell$ and ${\cal L}_{r,s}$, the corresponding 
separable state satisfies 
\begin{equation}
\label{theorem-3-2}
\mu_-^{\Gamma} = -\frac{1}{2} |q_1|,      \hspace{1.0cm}  \nu_-^{\Gamma} = \frac{1}{2} (1 - |\lambda_2 - \lambda_3|),
\end{equation}
where $\mu_-^{\Gamma}$ and $\nu_-^{\Gamma}$ are defined at Table II. Therefore, the CSS condition (\ref{deform-2}) implies
$q_1 = 0$, which results in ${\bm \tau}_{vp} = (0, 0, 1)$. If, therefore, this theorem is correct, the CSS of 
$\rho_{vp}$ is 
\begin{eqnarray}
\label{theorem-3-3}
\pi_{vp}&=& \frac{1}{4} \left[ I \otimes I + (\lambda_2 - \lambda_3) \left(\sigma_3 \otimes I + I \otimes \sigma_3 \right)
              + \sigma_3 \otimes \sigma_3 \right]               \\   \nonumber
        &=&  \left(          \begin{array}{cccc}
		             \frac{\lambda_1}{2} + \lambda_2  &  0  &  0  &  0       \\
					 0  &  0  &  0  &  0                               \\
					 0  &  0  &  0  &  0                               \\
					 0  &  0  &  0  &  \frac{\lambda_1}{2} + \lambda_3
					          \end{array}                             \right).
\end{eqnarray}

In order to show that $\pi_{vp}$ in Eq.(\ref{theorem-3-3}) is really CSS of $\rho_{vp}$, it is convenient to define 
another edge state
\begin{eqnarray}
\label{theorem-3-4}
\tilde{\pi}_{vp} = (\openone \otimes \sigma_x) \pi_{vp} (\openone \otimes \sigma_x)^{\dagger}
=        \left(         \begin{array}{cccc}
           \epsilon  & 0  &  0  &  0                                      \\
		   0  &  \frac{\lambda_1}{2} + \lambda_2  &  \epsilon  &  0       \\
           0  &  \epsilon  &  \frac{\lambda_1}{2} + \lambda_3  &  0       \\
		   0  &  0  &  0  &  \epsilon
		          \end{array}                                        \right),
\end{eqnarray}
where the infinitesimal positive parameter $\epsilon$ is introduced for convenience. This parameter will be 
taken to be zero after calculation. 

Let us define a edge state
\begin{eqnarray}
\label{theorem-3-5}
\sigma_Z = \left(               \begin{array}{cccc}
                       R_1  &  0  &  0  &  0               \\
					   0  &  R_2  &  Y  &  0               \\
					   0  &  Y  &  R_3  &  0               \\
					   0  &  0  &  0  &  R_4
					            \end{array}               \right)
\end{eqnarray}
with $Y = \sqrt{R_1 R_4}$ and $R_2 R_3 \geq R_1 R_4$. Then, by making use of Eq.(\ref{one-parameter-1-1})
Ref.\cite{miran-08-1} has shown that the set of the entangled states, which have $\sigma_Z$ as CSS, is represented
as 
\begin{eqnarray}
\label{theorem-3-6}
\rho_Z (x) = \left(           \begin{array}{cccc}
                     R_1 - x \bar{R}_1  &  0  &  0  &  0                    \\
					 0  &  R_2 - x \bar{R}_2  &  Y - x \bar{Y}  &  0        \\
					 0  &  Y - x \bar{Y}  &  R_3 - x \bar{R}_3  &  0        \\
					 0  &  0  &  0  &  R_4 - x \bar{R}_4
					           \end{array}                \right)
\end{eqnarray}
where $x \geq 0$ and\footnote{We corrected the sign mistake of Ref.\cite{miran-08-1}} 
\begin{eqnarray}
\label{theorem-3-7}
& &	\bar{R}_1 = \bar{R}_4 = \frac{Y^2}{R_1 + R_4}   \hspace{1.0cm}
	\bar{R}_2 = 2 Y^2 d \left[(R_2 - R_3) (R_2 L -z) + 2 Y^2 L \right]         \\   \nonumber
& & \bar{R}_3 = -2 	\bar{R}_1 - \bar{R}_2  \hspace{1.0cm}
    \bar{Y} = Y d \left[2 Y^2 (R_2 + R_3) L + (R_2 - R_3)^2 z \right].
\end{eqnarray}
In Eq.(\ref{theorem-3-7}) we define
\begin{equation}
\label{theorem-3-8}
z = \sqrt{(R_2 - R_3)^2 + 4 R_1 R_4}  \hspace{.7cm}
L = \ln \left(\frac{(R_2 + R_3) + z}{(R_2 + R_3) - z}  \right) \hspace{.7cm}
d = -\frac{1}{(R_1 + R_4) z^2 L}.
\end{equation}

Now, we identify $\sigma_Z$ with $\tilde{\pi}_{vp}$ by putting $R_1 = R_4 = Y = \epsilon$,
$R_2 = \lambda_1 / 2 + \lambda_2$ and $R_3 = \lambda_1 / 2 + \lambda_3$. Then, it is straightforward to 
compute $\bar{R}_1$, $\bar{R}_2$, $\bar{R}_3$, $\bar{R}_4$ and $\bar{Y}$. After taking $\epsilon \rightarrow 0$ 
limit, one can show $\bar{R}_1 = \bar{R}_2 = \bar{R}_3 = \bar{R}_4 = 0$ and 
$\bar{Y} = -|\lambda_2 - \lambda_3| / 2 L$, where
\begin{equation}
\label{theorem-3-9}
L = \ln \frac{1 + |\lambda_2 - \lambda_3|}{1 - |\lambda_2 - \lambda_3|} \geq 0.
\end{equation}
Therefore, the set of the entangled states, which have $\tilde{\pi}_{vp}$ as CSS, can be represented by 
\begin{eqnarray}
\label{theorem-3-10}
\tilde{\rho}_Z (x) = \left(              \begin{array}{cccc}
                            0  &  0  &  0  &  0                                                                     \\
							0  &  \frac{\lambda_1}{2} + \lambda_2  &  x \frac{|\lambda_2 - \lambda_3|}{2L}  &  0   \\
                            0  &  x \frac{|\lambda_2 - \lambda_3|}{2L}  &  \frac{\lambda_1}{2} + \lambda_3  &  0   \\
							0  &  0  &  0  &  0
							              \end{array}                       \right).
\end{eqnarray}
Finally, the set of the entangled states, which have $\pi_{vp}$ as CSS, can be derived by taking the inverse LU 
transformation, i.e.
\begin{equation}
\label{theorem-3-11}
\rho_Z (x) = (\openone \otimes \sigma_x)^{\dagger} \tilde{\rho}_Z (x) (\openone \otimes \sigma_x).
\end{equation}
It is easy to show that $\rho_Z (x)$ reduces to $\rho_{vp}$ in Eq.(\ref{VP-1}) when 
$x = x_{vp} = \lambda_1 L / |\lambda_2 - \lambda_3| \geq 0$, which completes the proof.

%%%%%%%%%%%%%%%%%%%%%%%%%%%%%%%%%%%%%%%%%%%%%%%%%%%%%%%%%
\begin{figure}[ht!]
\begin{center}
%\includegraphics[height=8cm]{fig5a.eps}
%\hspace{1.0cm}
%\includegraphics[height=8cm]{fig5b.eps}
\caption[fig5]{Fig. 5 shows how to find the CSS for the VP states. Fig. 5a and Fig. 5b correspond to 
$r=s=0.3$ and $r=s=0.5$, respectively. To find a CSS make a straight line $\ell$ first, which connects the 
nearest vertex of ${\cal T}$ and a point ${\bm t}_{vp}$. Second, compute the coordinate for the intersection point 
between $\ell$ and ${\cal L}_{r,s}$. Thirdly, identify the crossing point with ${\bm \tau}_{vp}$. Keeping 
${\bm u}_{vp} = {\bm r}_{vp}$ and ${\bm v}_{vp} = {\bm s}_{vp}$, one can find the CSS of the VP state.}
\end{center}
\end{figure}
%%%%%%%%%%%%%%%%%%%%%%%%%%%%%%%%%%%%%%%%%%%%%%%%%%%%%%%%%%%

Fig. 5 shows how to find the CSS for the VP state geometrically when $r=s=0.3$ (Fig. 5a) and $r=s=0.5$ (Fig. 5b). 
Fig. 5 indicates that the generalized VP states are on the edges of ${\cal T}$. First we make a line, 
which connects the nearest vertex of ${\cal T}$ and $\rho_{vp}$. 
Then, we compute the coordinate of the crossing point ${\bm \tau}$
between the line and ${\cal L}_{r,s}$. Finally, the CSS $\pi_{vp}$ of $\rho_{vp}$ can be computed 
by Eq.(\ref{VP-3}) with keeping the Bloch vectors.

\section{CSS for the Generalized Horodecki States}
In this section we discuss how to derive the CSS of the generalized Horodecki states. 
The Horodecki states are defined 
as mixture of one Bell state and separable states, which are orthogonal to the Bell state. One but most general 
example of the VP state is 
\begin{equation}
\label{H-1}
\rho_{H} = \lambda_1 \ket{\beta_1} \bra{\beta_1} + \lambda_2 \ket{01} \bra{01} + \lambda_3 \ket{10} \bra{10}
\end{equation}
with $\lambda_1 + \lambda_2 + \lambda_3 = 1$. 
By contrast with the VP state Horodecki state (\ref{H-1}) is separable when $\lambda_1^2 \leq 4 \lambda_2 \lambda_3$. 
This can be easily understood by computing the concurrence of $\rho_H$, which is 
\begin{eqnarray}
\label{concur-H}
{\cal C} (\rho_H) = \left\{             \begin{array}{cc}
                     \lambda_1 - 2 \sqrt{\lambda_2 \lambda_3}   
					 &  \hspace{1.0cm} \mbox{if} \hspace{.3cm} \lambda_1 \geq 2 \sqrt{\lambda_2 \lambda_3}    \\
                     0  &  \hspace{1.0cm} \mbox{if} \hspace{.3cm} \lambda_1 \leq 2 \sqrt{\lambda_2 \lambda_3}. 
					              \end{array}                \right.
\end{eqnarray}
Thus, ${\cal C} (\rho_H)$ becomes zero when $\lambda_1^2 \leq 4 \lambda_2 \lambda_3$, which indicates that 
$\rho_H$ is separable in this region.

Let the arbitrary Horodecki state be 
\begin{equation}
\label{H-2}
\rho_{H} = \frac{1}{4} \left( I \otimes I + {\bm r}_{H} \cdot {\bm \sigma} \otimes I + 
I \otimes {\bm s}_{H} \cdot {\bm \sigma} + \sum_{n=1}^3 (t_{H})_n \sigma_n \otimes \sigma_n \right)
\end{equation}
and its CSS be 
\begin{equation}
\label{H-3}
\pi_{H} = \frac{1}{4} \left( I \otimes I + {\bm u}_{H} \cdot {\bm \sigma} \otimes I + 
I \otimes {\bm v}_{H} \cdot {\bm \sigma} + \sum_{n=1}^3 (\tau_{H})_n \sigma_n \otimes \sigma_n \right).
\end{equation}
The following theorem shows how to compute ${\bm u}_{H}$, ${\bm v}_{H}$ and ${\bm \tau}_{H}$
from $\rho_{H}$. 

{\bf Theorem 4.} {\it If $\pi_{H}$ is a CSS of $\rho_{H}$, ${\bm u}_{H} = {\bm r}_{H}$ and 
${\bm v}_{H} = {\bm s}_{H}$. Let $\ell$ be a straight line, which connects ${\bm t}_{H}$ and the 
nearest vertex of ${\cal T}$. Then, ${\bm \tau}_{H}$ is the nearest 
crossing point between $\ell$ and ${\cal L}_{r,s}$.} 

\smallskip

{\bf Proof.} We will prove this theorem by following the same procedure of Theorem 3. Since other Horodecki states 
can be derived from $\rho_H$ in Eq.(\ref{H-1}) by LU transformation, it is sufficient to show that the CSS of 
Eq.(\ref{H-1}) satisfies this theorem. By identifying Eq.(\ref{H-1}) with Eq.(\ref{H-2}) one can easily show 
that ${\bm r}_H$, ${\bm s}_H$ and ${\bm t}_H$ become ${\bm r}_H = (0, 0, r)$, ${\bm s}_H = (0, 0, s)$ 
and ${\bm t}_H = (t_1, t_2, t_3)$, where
\begin{equation}
\label{theorem-4-1}
r = -s = \lambda_2 - \lambda_3  \hspace{1.0cm} t_1 = -t_2 = \lambda_1  \hspace{1.0cm} t_3 = 2 \lambda_1 - 1.
\end{equation}
Then, a point $P(q_1, q_2, q_3)$ on the line $\ell$ satisfies $q_2 = q_1$ and $q_3 = 2 q_1 - 1$. 

Let the point $P$ be crossing point between $\ell$ and ${\cal L}_{r,s}$. Then, $\mu_-^{\Gamma}$ and 
$\nu_-^{\Gamma}$ for the state corresponding to the point $P$ are given by
\begin{equation}
\label{theorem-4-2}
\mu_-^{\Gamma} = \frac{1}{2} \left[ (1 - q_1) - \sqrt{q_1^2 + (\lambda_2 - \lambda_3)^2} \right]   
\hspace{1.0cm}
\nu_-^{\Gamma} = \frac{q_1}{2}.
\end{equation}
Therefore, the CSS condition $\min (\mu_-^{\Gamma}, \nu_-^{\Gamma}) = 0$ gives two solutions
$P_1 = (q_1, -q_1, 2q_1 - 1)$ and $P_2 = (0, 0, -1)$, where
\begin{equation}
\label{heorem-4-3}
q_1 = \frac{1}{2} (\lambda_1 + 2 \lambda_2) (\lambda_1 + 2 \lambda_3).
\end{equation}
Since we have to choose the nearest point from ${\bm t}_H$, the solution we want is the former. Therefore, 
${\bm \tau}_H$ becomes $(q_1, -q_1, 2q_1 - 1)$. Then theorem 4 claims that the CSS of $\rho_H$ is 
\begin{eqnarray}
\label{theorem-4-4}
\pi_H&=&\frac{1}{4} \bigg[ I \otimes I + (\lambda_2 - \lambda_3) 
                                            \left\{\sigma_3 \otimes I - I \otimes \sigma_3 \right\}        \\  \nonumber
& &\hspace{4.0cm}        + q_1 \sigma_1 \otimes \sigma_1 - q_1 \sigma_2 \otimes \sigma_2 + (2q-1) \sigma_3 \otimes \sigma_3
		                                                                     \bigg]                       \\   \nonumber
     &=& \frac{1}{4} \left(              \begin{array}{cccc}
(\lambda_1 + 2\lambda_2)(\lambda_1 + 2\lambda_3)  &  0  &  0  &  (\lambda_1 + 2\lambda_2)(\lambda_1 + 2\lambda_3) 
                                                                                                       \\
0  &  (\lambda_1 + 2\lambda_2)^2  &  0  &  0                                                          \\
0  &  0  &  (\lambda_1 + 2\lambda_3)^2  &  0                                                          \\
(\lambda_1 + 2\lambda_2)(\lambda_1 + 2\lambda_3)  &  0  &  0  &  (\lambda_1 + 2\lambda_2)(\lambda_1 + 2\lambda_3)
                                           \end{array}               \right).
\end{eqnarray}
														
Now, we will show that $\pi_H$ in Eq.(\ref{theorem-4-4}) is really CSS of $\rho_H$ by making use of 
Eq.(\ref{one-parameter-1-1}). In order to show this we define 
$\tilde{\pi}_H = (\openone \otimes \sigma_x) \pi_H (\openone \otimes \sigma_x)^{\dagger}$. Then by making use of 
Eq.(\ref{theorem-3-5}) and Eq.(\ref{theorem-3-6}) it is straightforward
to find a set of the entangled quantum states $\tilde{\rho} (x)$, whose CSS are $\tilde{\pi}_H$. After taking the inverse
LU transformation one can derive 
$\rho (x) = (\openone \otimes \sigma_x)^{\dagger} \tilde{\rho} (x) (\openone \otimes \sigma_x)$. The expression 
of $\rho (x)$ is 
\begin{eqnarray}
\label{theorem-4-5}
\rho (x) = \left(                   \begin{array}{cccc}
                  Y + x \eta  &  0  &  0  &  Y + x \eta                      \\
				  0  &  R_1 - x \eta  &  0  &  0                             \\
				  0  &  0  &  R_4 - x \eta  &  0                             \\
				  Y + x \eta  &  0  &  0  &  Y + x \eta 
				                      \end{array}               \right)
\end{eqnarray}
where $x \geq 0$ and 
\begin{eqnarray}
\label{theorem-4-6}
& & R_1 = \frac{1}{4} (\lambda_1 + 2 \lambda_2)^2    \hspace{1.0cm}
    R_4 = \frac{1}{4} (\lambda_1 + 2 \lambda_3)^2                         \\   \nonumber
& & Y = \frac{1}{4} (\lambda_1 + 2 \lambda_2) (\lambda_1 + 2 \lambda_3)   \hspace{1.0cm}
    \eta = \frac{Y^2}{R_1 + R_4}.
\end{eqnarray}
When 
\begin{equation}
\label{theorem-4-7}
x = x_H = \frac{1}{\eta} \left(\frac{\lambda_1}{2} - Y \right),
\end{equation}
$\rho (x)$ reduces to $\rho_H$ in Eq.(\ref{H-1}). It is easy to prove that $x_H \geq 0$ if 
$\lambda_1^2 \geq 4 \lambda_2 \lambda_3$, which is an entangled condition for $\rho_H$. Therefore, 
theorem 4 is completely proved. 					

%%%%%%%%%%%%%%%%%%%%%%%%%%%%%%%%%%%%%%%%%%%%%%%%%%%%%%%%%
\begin{figure}[ht!]
\begin{center}
\caption[fig6]{Fig. 6 shows how to find the CSS for the generalized Horodecki states. Fig. 6a and Fig. 6b correspond to 
$r=-s=0.3$ and $r=-s=0.5$ respectively. In order to find CSS one makes a straight line $\ell$ first, which connects the 
nearest vertex of ${\cal T}$ and a point ${\bm t}_{H}$. Secondly, we compute the coordinate for the nearest crossing point 
between $\ell$ and ${\cal L}_{r,s}$. Thirdly, we identify the crossing point with ${\bm \tau}_{H}$. Keeping 
${\bm u}_{H} = {\bm r}_{H}$ and ${\bm v}_{H} = {\bm s}_{H}$, one can find the CSS of the Horodecki state.}
\end{center}
\end{figure}
%%%%%%%%%%%%%%%%%%%%%%%%%%%%%%%%%%%%%%%%%%%%%%%%%%%%%%%%%%%

Fig. 6 shows how to find the CSS for the generalized Horodecki state $\rho_H$ 
when $r=-s= 0.3$ (Fig. 6a) and $r=-s=0.5$ (Fig. 6b). 
If $\rho_H$ is explicitly given, compute ${\bm r}_H$, ${\bm s}_H$ and ${\bm t}_H$. Then make a straight 
line which connects a point ${\bm t}_H$ and the nearest vertex of ${\cal T}$. Find the crossing points between the 
line and ${\cal L}_{r,s}$. As Fig. 3 shows, there are two intersection points $P_1$ and $P_2$ for the Horodecki states. This 
is why the CSS condition $\min (\mu_-^{\Gamma}, \nu_-^{\Gamma}) = 0$ gives two different solutions. 
Using the nearest crossing point ($P_1$ in Fig. 6) one can derive ${\bm \tau}_H$ straightforwardly. Finally using 
Eq.(\ref{H-3}) with imposing ${\bm u}_H = {\bm r}_H$ and ${\bm v}_H = {\bm s}_H$, one can derive
$\pi_H$, the CSS of $\rho_H$.

\section{Difficulties in finding CSS for arbitrary states}
In the previous sections we have shown how to find the CSS for the Bell-diagonal, generalized VP, and generalized Horodecki
states. In fact, it is possible to find the CSS because those states exhibit the following nice features. Let ${\bm r}$, 
${\bm s}$ and ${\bm t}$ be Bloch and correlation vectors of those states. Let ${\bm u}$, ${\bm v}$ and ${\bm \tau}$ be 
Bloch and correlation vectors for the corresponding CSS of those states. Then, the features are:

\noindent
(i) ${\bm u} = {\bm r}$ and ${\bm v} = {\bm s}$. 

\noindent
(ii) ${\bm \tau}$ can be computed from the crossing point between the straight line $\ell$ and the surface for a set of 
the separable states ${\cal L}_{r,s}$.

However, such simple but nice features are not maintained for the general mixtures. For example, let us consider the 
comparatively simple model introduced in Eq.(\ref{theorem-3-5}) and Eq.(\ref{theorem-3-6}). It is straightforward to show 
that the first property, i.e. ${\bm u} = {\bm r}$ and ${\bm v} = {\bm s}$, is not maintained unless $R_2 = R_3$\footnote{When 
$R_2 = R_3$, one can show $[\rho_Z(x), \sigma_Z] = 0$. Therefore, 
$E_R (\rho_Z(x) \otimes \tilde{\rho}) = E_R (\rho_Z(x)) + E_R (\tilde{\rho})$ for all two-qubit mixture 
$\tilde{\rho}$\cite{rains-99-2}.}. 
In order to find the CSS for the arbitrary states, therefore, we have to find the
explicit relations between $({\bm r}, {\bm s})$ and $({\bm u}, {\bm v})$. As far as we know, still this is an unsolved problem.

%%%%%%%%%%%%%%%%%%%%%%%%%%%%%%%%%%%%%%%%%%%%%%%%%%%%%%%%%
\begin{figure}[ht!]
\begin{center}
\includegraphics[height=7cm]{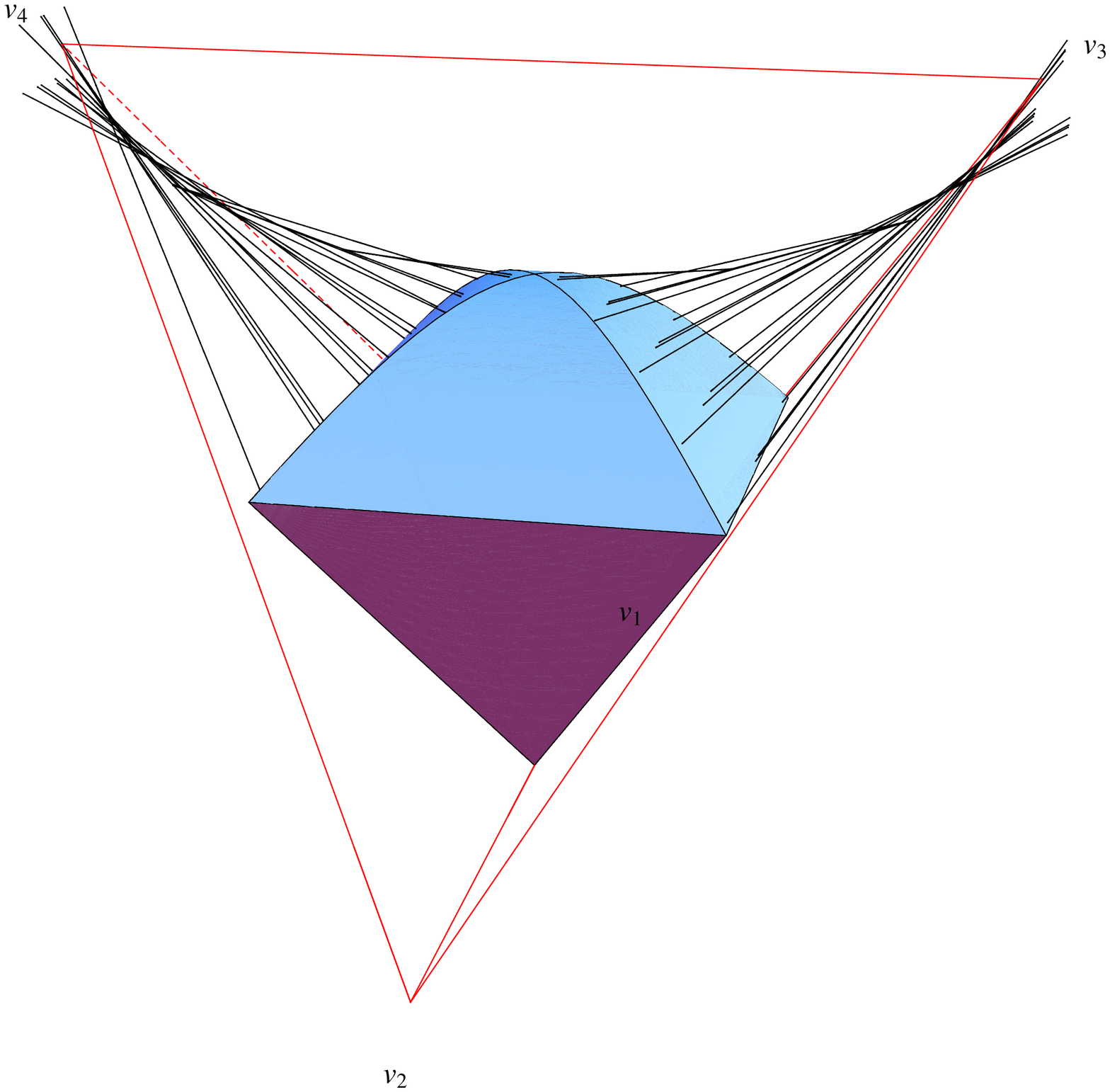}
\includegraphics[height=7cm]{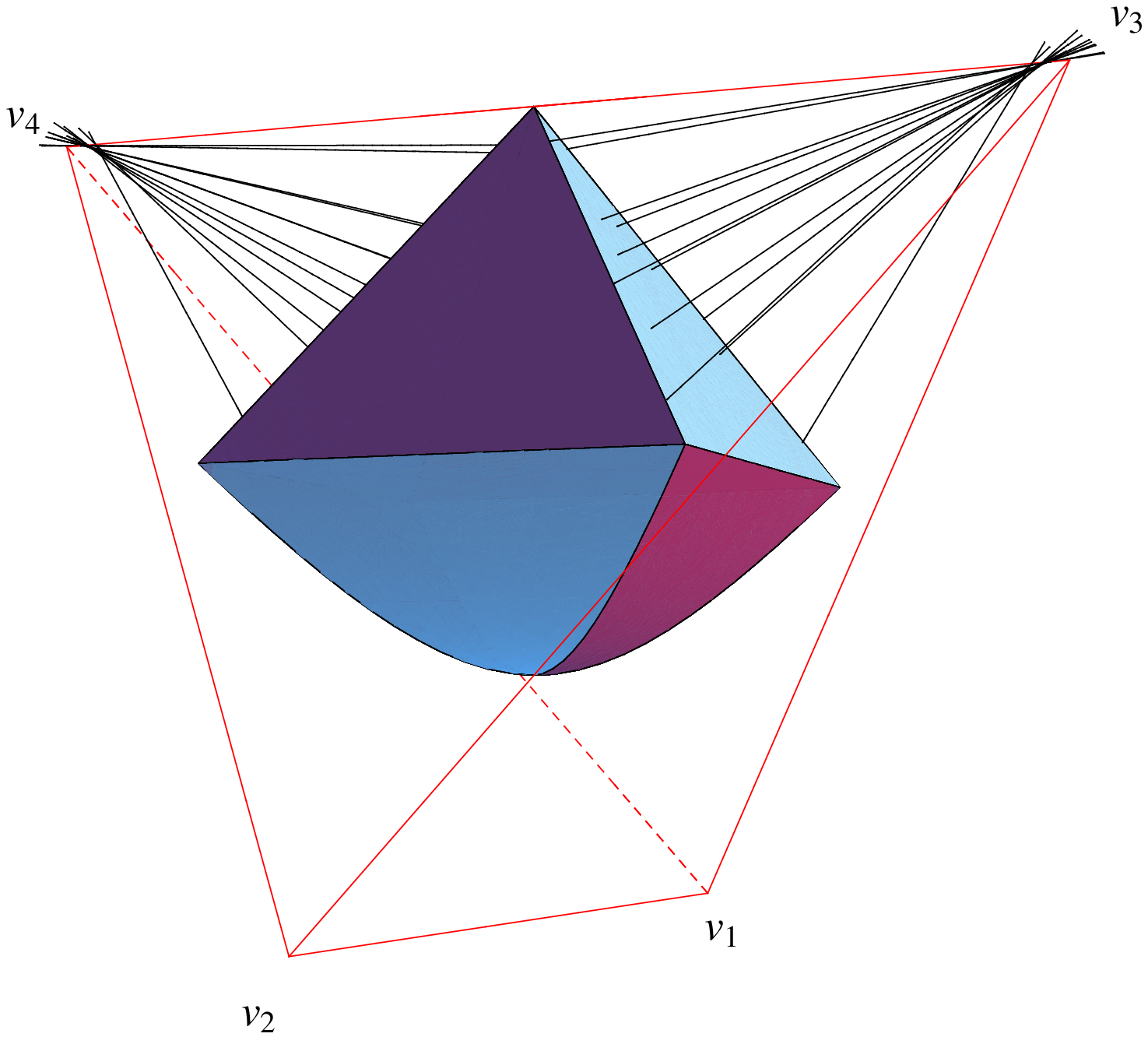}

\includegraphics[height=7cm]{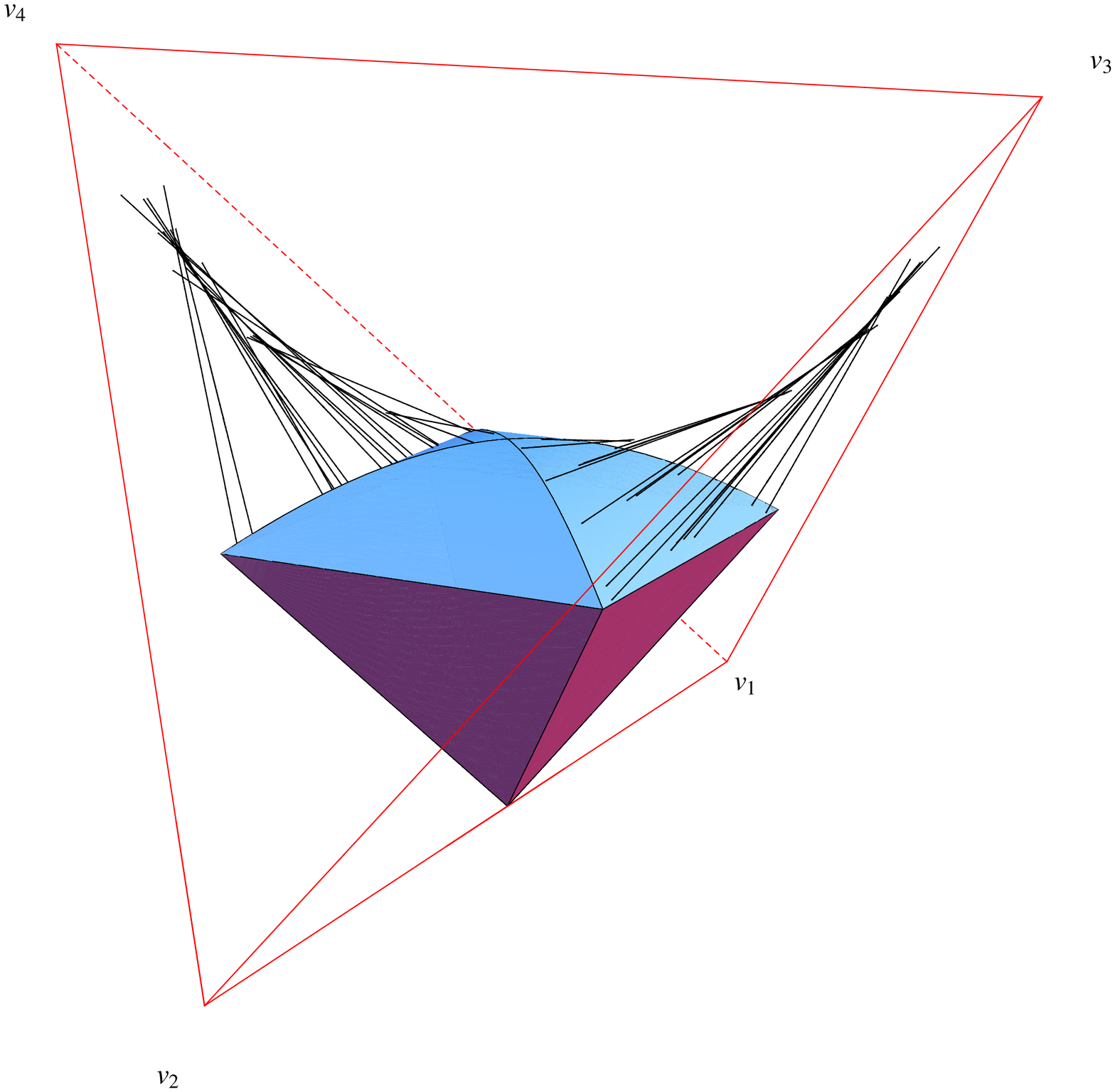}
\includegraphics[height=7cm]{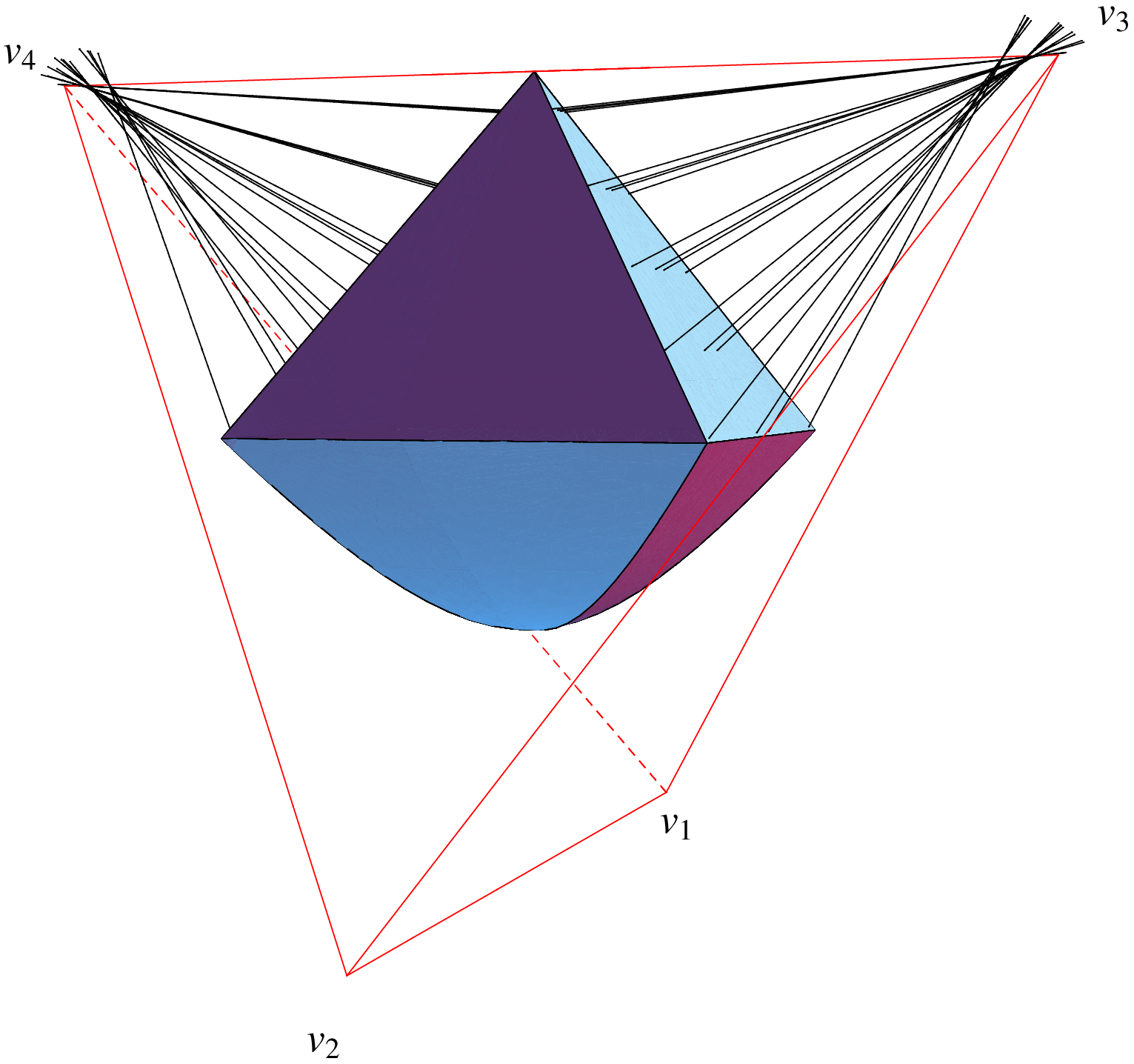}
\caption[fig7]{Fig. 7 shows that the property (ii) is not maintained for the arbitrary two-qubit states. 
Four figures in Fig. 7 correspond to $r=s=0.3$ (Fig. 7a), 
$r=-s=0.3$ (Fig. 7b), $r=s=0.5$ (Fig. 7c) and $r=-s=0.5$ (Fig. 7d). For convenience, we plot ${\cal T}$ and 
${\cal L}_{r,s}$ together in each figure. Each line in Fig. 7 represents a set of ${\bm t}$, whose CSS has same 
${\bm \tau}$. As Fig. 7 has exhibited, not all lines pass one of the vertices of ${\cal T}$. This fact 
indicates that the nice property (ii) is not maintained for the general mixtures.}
\end{center}
\end{figure}
%%%%%%%%%%%%%%%%%%%%%%%%%%%%%%%%%%%%%%%%%%%%%%%%%%%%%%%%%%%

In addition, one can show that the second property is not maintained too for the general mixtures.
Using Eq.(\ref{one-parameter-1-1}) we plot the ${\bm \tau}$ (correlation vector of CSS)-dependence of 
${\bm t}$ (correlation vector of the entangled state) with varying the parameter $x$
when $\min (\mu_-^{\Gamma}, \nu_-^{\Gamma}) = \nu_-^{\Gamma} = 0$.
Since similar behavior arises when $\min (\mu_-^{\Gamma}, \nu_-^{\Gamma}) = \mu_-^{\Gamma} = 0$, 
we have not included this case in Fig. 7. Four figures in Fig. 7 correspond to, respectively, $r=s=0.3$ (Fig. 7a), 
$r=-s=0.3$ (Fig. 7b), $r=s=0.5$ (Fig. 7c) and $r=-s=0.5$ (Fig. 7d). For convenience, we plot ${\cal T}$ and 
${\cal L}_{r,s}$ together in each figure. Each line in Fig. 7 represents a set of ${\bm t}$, whose CSS has same 
${\bm \tau}$. As Fig. 7 exhibits, not all lines do pass one of the vertices of ${\cal T}$. This fact 
indicates that unfortunately the nice property (ii) is not maintained for the arbitrary states.

The non-maintenance of the property (ii) can be proved on the analytical ground by making use of the simpler 
model. Let us consider $\sigma_Z$ in Eq.(\ref{theorem-3-5}) and $\rho_Z(x)$ in Eq.(\ref{theorem-3-6}). Then, the 
Bloch vectors ${\bm r}$, ${\bm s}$ and the correlation vector ${\bm t}$ of $\rho_Z(x)$ are 
${\bm r} = (0, 0, r)$, ${\bm s} = (0, 0, s)$ and ${\bm t} = (t_1, t_2, t_3)$, where
\begin{eqnarray}
\label{property2-1}
& &r = (R_1 + R_2 - R_3 - R_4) - x (\bar{R}_2 - \bar{R}_3)    \\   \nonumber
& &   s = (R_1 - R_2 + R_3 - R_4) + x (\bar{R}_2 - \bar{R}_3)    \\   \nonumber
& &t_1 = t_2 = 2 Y - 2 x \bar{Y}                              \hspace{.5cm}
   t_3 = (R_1 - R_2 - R_3 + R_4) -  4 x \bar{R}_1.
\end{eqnarray}
Of course, if we take $x \rightarrow 0$ limit in Eq.(\ref{property2-1}), the corresponding quantities are 
the Bloch vectors and the correlation vector of $\sigma_Z$. Now, let us consider another state $\pi_Z$, 
which can be obtained from $\sigma_Z$ by changing $Y \rightarrow Y'$ and $R_i \rightarrow R_i' (i=1, \cdots , 4)$.
In order to ensure that $\pi_Z$ is CSS we require $Y' = \sqrt{R_1' R_4'}$ and $R_2' R_3' \geq R_1' R_4'$.
Then, the set of the entangled states $\xi_Z (x')$, whose CSS are $\pi_Z$, can be obtained from $\rho_Z(x)$ by 
changing $Y \rightarrow Y'$, $R_i \rightarrow R_i' (i=1, \cdots , 4)$ and $x \rightarrow x'$. Thus, Bloch vectors
${\bm r'}$, ${\bm s'}$ and correlation vector ${\bm t'}$ of $\xi_Z (x')$ are 
${\bm r'} = (0, 0, r')$, ${\bm s'} = (0, 0, s')$ and ${\bm t'} = (t_1', t_2', t_3')$, where
\begin{eqnarray}
\label{property2-2}
& &r' = (R_1' + R_2' - R_3' - R_4') - x' (\bar{R}_2' - \bar{R}_3')    \\   \nonumber
& &   s' = (R_1' - R_2' + R_3' - R_4') + x' (\bar{R}_2' - \bar{R}_3')    \\   \nonumber
& &t_1' = t_2' = 2 Y' - 2 x' \bar{Y}'                              \hspace{.5cm}
   t_3' = (R_1' - R_2' - R_3' + R_4') -  4 x' \bar{R}_1'.
\end{eqnarray}
Then, it is straightforward to show that the condition ${\bm t} = {\bm t'}$ imposes
\begin{equation}
\label{property-2-3}
x = \frac{\bar{Y}' (\tilde{r}-\tilde{r}') - 4 (Y - Y') \bar{R}_1'}{4 (\bar{Y}' \bar{R}_1 - \bar{Y} \bar{R}_1')}
                                                                            \hspace{.5cm}
x' = \frac{\bar{Y} (\tilde{r}-\tilde{r}') - 4 (Y - Y') \bar{R}_1}{4 (\bar{Y}' \bar{R}_1 - \bar{Y} \bar{R}_1')}
\end{equation}
where $\tilde{r} = R_1 - R_2 - R_3 + R_4$ and $\tilde{r}' = R_1' - R_2' - R_3' + R_4'$. Thus, one can compute the 
crossing point ${\bm t} = {\bm t'} = (\mu_1, \mu_2, \mu_3)$, where $\mu_i$ becomes
\begin{eqnarray}
\label{property-2-4}
& &\mu_1 = \mu_2 = \frac{4 (Y \bar{Y}' \bar{R}_1 - Y' \bar{Y} \bar{R}_1') - \bar{Y} \bar{Y}' (\tilde{r} - \tilde{r}')}
                      {2 (\bar{Y}' \bar{R}_1 - \bar{Y} \bar{R}_1')}             \\   \nonumber
& &\mu_3 = \frac{4 (Y - Y') \bar{R}_1 \bar{R}_1' - (\tilde{r} \bar{Y} \bar{R}_1' - \tilde{r}' \bar{Y}' \bar{R}_1)}
              {\bar{Y}' \bar{R}_1 - \bar{Y} \bar{R}_1'}.
\end{eqnarray}

As a special case we consider the Bell-diagonal case by letting $R_1 = R_4 = Y = 2 \bar{R}_1 = 2\bar{R}_4 = -2\bar{R}_2 = -2 \bar{R}_3 = a$,
$R_2 = R_3 = -2 \bar{Y} = b$, $R_1' = R_4' = Y' = 2 \bar{R}_1' = 2\bar{R}_4' = -2\bar{R}_2' = -2 \bar{R}_3' = a'$ and 
$R_2' = R_3' = -2 \bar{Y}' = b'$. Of course, one can show directly that $\rho_Z(x)$ and $\xi_Z(x')$ are really Bell-diagonal 
states. Using the normalization conditions $2(a+b) = 2 (a'+b') = 1$, it is easy to verify that the crossing point $(\mu_1,\mu_2,\mu_3)$
is simply $\mu_1=\mu_2=1$ and $\mu_3 = -1$, which is one of the vertices of ${\cal T}$. It is worthwhile noting that 
the crossing point is independent of particular choice of Bell-diagonal states $\rho_Z(x)$ and $\xi_Z(x')$. This fact implies that
all straight lines, which connect ${\bm \tau}$ and ${\bm t}$, pass one of the vertices of ${\cal T}$, which is consistent with
theorem 2. 

However for the arbitrary mixtures Eq.(\ref{property-2-4}) implies that the crossing point 
$(\mu_1, \mu_2, \mu_3)$ is dependent on the choice of the entangled states $\rho_Z(x)$ and $\xi_Z(x')$. This is why the nice 
property (ii), which holds for the Bell-diagonal, generalized VP, and generalized Horodecki states, does not hold for the arbitrary 
mixture as Fig. 7 has indicated. In order to, therefore, derive the closed formula of $E_R (\rho)$ for the arbitrary two-qubit mixtures
$\rho$, we have to understand how the property (ii) is modified when ${\bm r}$, ${\bm s}$ and ${\bm t}$ are arbitrary. 
Unfortunately, still this is an unsolved problem too. 

\section{Conclusion}
In this paper we have considered how to find the CSS in the two-qubit system from the geometrical point of view. 
Of course, one can straightforwardly compute
the REE of the state $\rho$ if its CSS is found. Therefore, it is important to develop a technique for finding CSS to overcome
the calculational difficulty of the REE. Since, furthermore, the REE is a tight upper bound of the distillable entanglement, finding 
CSS is also important to understand the nature of the optimal (or near-optimal) purification protocols. 

If $\rho$ is one of Bell-diagonal, generalized VP, and generalized Horodecki states, we have shown how to find the CSS of 
$\rho$, {\it say} $\sigma$, systematically by proving the following nice two properties: 
(i) The Bloch vectors of $\sigma$ are identical 
with those of $\rho$. (ii) The correlation vector of $\sigma$ exactly corresponds to the crossing point between
the line $\ell$ and the geometrical object ${\cal L}_{r,s}$. Using these two properties it is straightforward to find
the CSS of $\rho$. 

As we have shown in the previous section, however, these two nice properties are not maintained for the general two-qubit 
states. Therefore, in order to derive the closed formula of $E_R (\rho)$ for the arbitrary mixture $\rho$ we have to understand 
how these two properties are modified when Bloch and correlation vectors of $\rho$ are arbitrary. The research into these
issues is in progress and will be reported elsewhere.

Another interesting issue, which we will go further, is to explore the REE and the distillable entanglement for the 
higher-qubit or qudit systems. Few years ago, the analytical expressions of the distillable entanglement are obtained 
for some higher-dimensional bipartite states\cite{ghosh01,chen02,chen03}. Authors in those references used the upper-bound 
criterion $D \leq E_R$ and separability property of the Smolin's unlockable state\cite{smolin00} in various cuts.  
We would like to modify Eq.(\ref{one-parameter-1-1}) to be applicable not only for low-rank $\sigma^*$ but also for 
higher-dimensional system. If the generalization of Eq.(\ref{one-parameter-1-1}) is possible, we can use it to compute
the REE and the distillable entanglement for many more higher-dimensional states. It may enable us to understand the nature of 
the optimal (or near-optimal) purification protocols. This work is in progress too.

\begin{acknowledgments}
This work was supported by National Research Foundation of Korea Grant funded by the
Korean Government (2009-0073997).
\end{acknowledgments}

\end{document}